\documentclass{article} 
\usepackage{iclr2025_conference,times}
\usepackage{graphicx} 
\usepackage[a4paper, total={6in, 8in}]{geometry}
\usepackage{amsmath}
\usepackage{booktabs}
\usepackage{array}
\usepackage[table]{xcolor}
\usepackage{colortbl}
\usepackage{hhline}
\usepackage{multirow}
\usepackage{rotating}    
\usepackage{ragged2e}
\usepackage{diagbox}
\usepackage{makecell}  

\usepackage{amsmath,amsfonts,bm}

\def\eqref#1{equation~\ref{#1}}

\def\1{\bm{1}}

\DeclareMathAlphabet{\mathsfit}{\encodingdefault}{\sfdefault}{m}{sl}
\SetMathAlphabet{\mathsfit}{bold}{\encodingdefault}{\sfdefault}{bx}{n}

\usepackage{hyperref}
\usepackage{url}

\iclrfinalcopy
\title{\textbf{Watermarking Large Language Models in Europe}:
\\ Interpreting the AI Act in Light of Technology
}

\author{\small Thomas Souverain* \\ Postdoctoral Fellow, Department of AI Ethics, CEA Paris-Saclay, France. \\ * E-mail: \url{thomas.souverain@cea.fr}}

\begin{document}

\maketitle

\begin{abstract}

To foster trustworthy Artificial Intelligence (AI) within the European Union, the AI Act requires providers to mark and detect the outputs of their general-purpose models.
The Article 50 and Recital 133 call for marking methods that are "sufficiently reliable, interoperable, effective and robust".
Yet, the rapidly evolving and heterogeneous landscape of watermarks for Large Language Models (LLMs) makes it difficult to determine how these four standards can be translated into concrete and measurable evaluations. 
Our paper addresses this challenge, anchoring the normativity of European requirements in the multiplicity of watermarking techniques. Introducing clear and distinct concepts on LLM watermarking, our contribution is threefold:

\textbf{Watermarking Categorisation: }
We propose an accessible taxonomy
of watermarking methods according to the stage of the LLM lifecycle at which they are applied — before, during, or after training, and during next-token distribution or sampling.

\textbf{Watermarking Evaluation: }
We interpret the EU AI Act’s requirements by mapping each criterion with state-of-the-art evaluations on robustness and detectability of the watermark, and of quality of the LLM. 
Since interoperability remains largely untheorised in LLM watermarking research, we propose three normative dimensions to frame its assessment. 

\textbf{Watermarking Comparison: }
We compare current watermarking methods for LLMs against the operationalised European criteria and show that no approach yet satisfies all four standards. Encouraged by emerging empirical tests, we recommend further research into watermarking directly embedded within the low-level architecture of LLMs. 
    
\end{abstract}

\section*{\centering Keywords}
\begin{center}
    Large Language Models (LLMs), Watermarks, Taxonomy, Europe, Normativity, Operational.
\end{center}

\section{Introduction}\label{sec:intro_eu_ethical_and_market_needs}

Among the most advanced techniques in Artificial Intelligence (AI), Large Language Models (LLMs) can generate text that appears meaningful, though it is not strictly derived from human programming or explicit instructions. The capabilities of LLMs therefore call for watermarking or digital signatures to authenticate synthetic content \citep{Watermark_History_HandMadePaper}, to maintain the distinction between AI- and human-authored texts. Beyond copyright infringement, watermarking is crucial to prevent emotional and cognitive confusion, which may result in serious social and psychological harm \citep{grinbaum2022ethical}. 

These ethical foundations have been recently nailed into legal obligations. The European Union (EU) has positioned itself as a pioneer in promoting watermarking for generative models. The use of watermarks to identify text outputs produced by LLMs was first proposed in Opinion 7 of the French Committee for Digital Ethics \citep{CNPEN2023_avis7}, later discussed at the G7 meeting in Japan \citep{g7_2023_hiroshima_process_genAI}, and formally included in the AI Act \citep{AI_Act_Europe}. 
In particular, Article 50(2) and Recital 133 of the Act stipulate that LLM outputs should be marked by methods that are “sufficiently reliable, interoperable, effective and robust, as far as this is technically feasible” \citep{AI_Act_Europe}. However, neither the Act itself nor the accompanying Code of Practice \citep{Code_of_Practice_GPAI_Europe} provides specific guidance on how these standards relate to existing watermarking techniques for LLMs.


Our paper aims to bridge this gap between legal norms and technical realities by operationalizing the European requirements for LLM watermarking. We clarify how the AI Act’s four criteria apply to synthetic content authentication, grounding our analysis in a precise overview of watermarking methods for LLMs. Section \ref{sec:existing_techniques_presentation} outlines the main types of LLM watermarks, Section \ref{sec:evaluating_pros_contras} reviews existing evaluation practices, and Section \ref{sec:links_scient_eu_evaluation} offers an operational interpretation of the AI Act criteria. Section \ref{sec:eu_interoperability} develops research foundations for the underexplored criterion of interoperability. Thanks to our operational interpretation of EU criteria, Section \ref{sec:overall_strengths_weaknesses} finally trades off the advantages and weaknesses of existing LLM watermarks. 

 
 To guide the reader from European legal principles to technical feasibility in LLMs, we provide three concise visual summaries of our main results. Figure 1. presents a schematic overview of watermarking styles, situating them within the LLM lifecycle. Figure \ref{fig:overlap_eu_criteria} illustrates how operational interpretations of the AI Act’s requirements for watermarking are derived by resolving conceptual overlaps. Table \ref{tab:strengths_weaknesses_watermarks_llm_overview} offers a comparative overview of the advantages and limitations of the main watermarking families, assessed against the operationalized European criteria. Together, these visual aids clarify how LLM watermarking can be implemented and assessed for compliance.

\section{State of the art in LLM Watermarking}\label{sec:existing_techniques_presentation}

To assess whether the EU AI Act criteria match LLM watermarking techniques, we must present them in a clear manner. Therefore, we aim to be as concise as it may in formalism and details ; for each approach, we only select the most discussed and representative techniques.\footnote{We refer to \cite{liu2024_text_survey} and \cite{gloaguen2025_open_source} for more formalism, and to \cite{liang2024watermarking} and \cite{zhao2025_watermark_overview} for more exhaustive lists on papers that share similar approaches.}

From the observation that current classifications of watermarking techniques for LLMs either lack of clarity, precision or completeness (Section \ref{sec:our_taxonomy}), we introduce a simple and original taxonomy. We distinguish between methods that stamp tokens and texts outside LLM computations (Section \ref{sec:pre_post_processing_watermarking}), and marks inside the architecture and generation process of the model (Section \ref{sec:in_processing_watermarking_more_formal}). 

\subsection{Simplifying existing taxonomies}\label{sec:our_taxonomy}

Market and legal incentives carved out a plethora of LLM watermarking techniques. However \cite{Meta_taxonomy_fernandez} point out that emerging taxonomies lack of clarity to define and characterize these tree fern LLM watermarking methods. As the authors observe, even the broad picture of evaluation methods provided by \cite{zhao2025_watermark_overview} interchangeably use terms as "semantic" and "in-processing" methods. 
Other taxonomies on watermarking for neural networks as \cite{Boenisch_2021_multi_bits_nn_taxonomy} are based on watermarks' properties such as one-bit storage (only indicating if the content was watermarked) or multi-bit format (e.g. with further information on the provider, the generation time), though they do not directly relate these properties with stages of the model's development. 
\cite{Meta_taxonomy_fernandez} themselves lack of completeness to describe the diversity of in-processing and post-hoc LLM watermarking. 


- \cite{liang2024watermarking} fittingly demarcates between watermarks which are embedded "into text" and "into model". Nonetheless, it is unclear why the modification of the LLM generation process shall be tied to the "text" and not the "model" watermarking. Furthermore, by assigning watermarking "based on cryptography" its own category, the authors obscure the extent to which it is aligned with textual or model-oriented methods.


- \cite{DeepMind_dathathri_2024_complexifying_tokens} separate "retrieval-based approach", "post-hoc detection", and "text watermarking". While the purpose of their paper is to set up a new watermarking method, this introductory taxonomy seems far from obvious. The two first approaches they mention are exclusively dedicated to \textit{detect} watermarks, finding statistical patterns or training specific classifiers to identify LLM-generated passages. Notwithstanding, their third category of "text watermarking" involves detection techniques which can overlap with retrieval and post-hoc methods.  
This classification reveals a blurring of two dimensions: the stage in the LLM development cycle when the watermark manifests, and the procedural step of watermarking, whether concerning its design or its detection.

- In current LLM watermarking taxonomies, the most convincing we found was \cite{liu2024_text_survey}. The authors introduce a rare, yet decisive distinction during the generation time between logit generation, and token sampling. 
We incorporate this distinction into our taxonomy, and also draw inspiration from \cite{gloaguen2025_open_source} to characterize in-model watermarking techniques. Our taxonomy originally combines them, as \cite{liu2024_text_survey} incompletely analyze watermarking techniques inside LLMs, vaguely mentioning "triggers", or ignoring watermarking signals into LLM weights ; while \cite{gloaguen2025_open_source} narrowly study them in open source contexts and do not stress on the yet crucial notion of logits. 

Having noticed the non-exhaustive defaults of current taxonomies listed above, we propose here a simple and novel taxonomy to distinguish them according to a temporal criterion, that is, \textit{when} they take place in regard to the LLM generation process. 
As \cite{DeepMind_dathathri_2024_complexifying_tokens} and \cite{Meta_taxonomy_fernandez} highlight, the watermarking signal can be added before, during, or after the LLM generates any text.

\subsection{Pre- and Post-Processing Approaches: syntactic and semantic}\label{sec:pre_post_processing_watermarking}

While they differ in their object, being either pre-training data or LLM-generated text, we note that the watermarks taking place before and after LLM training and inference display similar styles. 

A first type of techniques is \textit{syntactic}. \cite{Wei_2024_pre_character_substitution_pseudo_random} offer a random substitution of Unicode characters. This watermarking signature is then embedded in the training data, feeding the LLM. Corresponding to this pre-processing method, character substitution also exists once the model has been trained. For instance, EASYMARK \citep{Sato_2023_post_character_subst_EASYMARK} applies the same idea to LLM-generated text, where hidden characters are concatenated.

Besides, pre- and post-processing watermarking allows for \textit{semantic} techniques. In that direction, \cite{Zhang_2024_post_nn_syntax_replacement} propose a subtle method to watermark the generated text through synonym-substitution. Based on the frequency of generated words, the authors generate context-based synonyms, purposed to be more present in the LLM contents. 

At the intersection of semantic and syntactic approaches, specific classifiers are trained to detect and replace parts of the data fueling LLMs or generated by them. \cite{Zhang_2024_post_nn_syntax_replacement} train a neural network to focus on punctuation or prepositions and propose discrete substitutions, while \cite{Abdelnabi_2021_post_decoder_adversarial_replacement} use an encoder / decoder couple to generate substitutions, minimizing their detectability by malicious actors in LLM contents.

\subsection{In-Processing Approaches: Training, weights, distribution and sampling}\label{sec:in_processing_watermarking_more_formal}

Whereas watermarking in training datasets or already-generated texts is often practical to implement and flexible to test \citep{liang2024watermarking}, inserting watermarking signatures directly into the LLM or its inference process has the potential to withstand quantization, pruning and fine-tuning more effectively \cite{gloaguen2025_open_source} and to be less detectable for attackers \cite{DeepMind_dathathri_2024_complexifying_tokens}. 

Given a vocabulary of tokens $V$, we can formalize a Language Model as a function $p_\theta : V^{(t-1)} \rightarrow \Delta(V) $, learning one's parameters $\theta$. $p_\theta$ maps the $(t-1)$ first tokens to a logits' distribution of tokens in $\Delta(V)$. As GPT, Claude, Gemini, Mistral and most LLM iteratively produce their text, each token being sampled conditionally on the others, we will focus here on autoregressive LLMs. That is, given a prompt $x$ and the already generated tokens $v_1, ..., v_{(t-1)}\in V$, the LLM selects the token $v_t$ with the highest conditional probability $p(.|x\cap v_1,..,v_{(t-1)})$. In our view, there are four main steps where watermarking can enter the LLM inference process. We present these stages here, synthetised in Figure 1. above:

\subsubsection*{Into LLM Architecture: steps 1 \& 2}

\begin{figure}[t]
\label{fig:watermarking_during_inference}
\caption{Watermarking a LLM by altering the Generation Process: Four Steps}
\centering
\resizebox{\textwidth}{!}{
\includegraphics{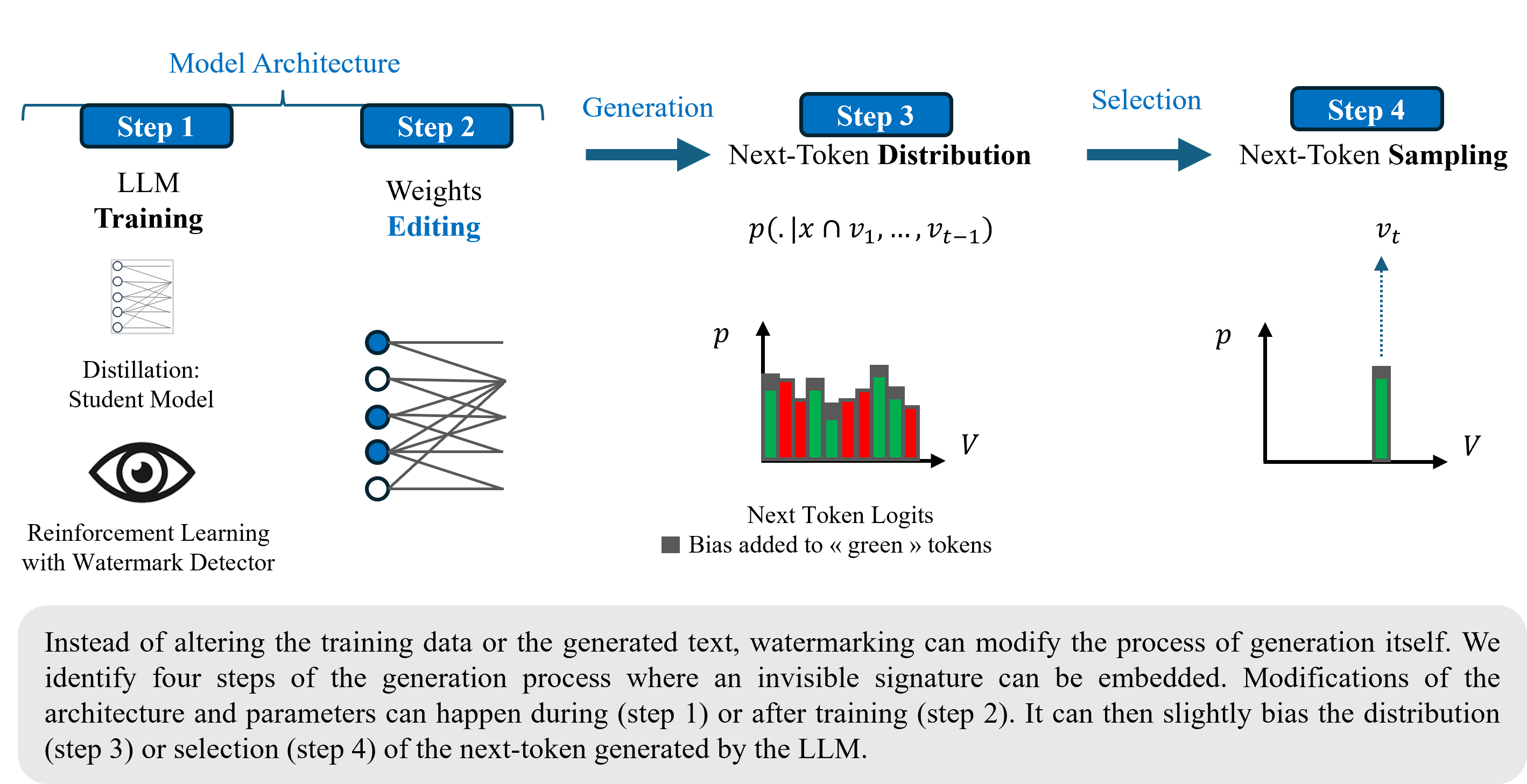}}
\end{figure} 

\subsubsection*{Step 1: LLM Training}\label{itm: In-Processing - Step 1}
The first kind of watermarking takes place during LLM training. The watermark directly changes the model's architecture. 
Some authors have successfully realized it through \textit{distillation} \citep{Gu_2023_distillation_watermarking}. The watermarking signal slightly biases a LLM generation (see Step 3 below). This LLM is then used as a teacher to train a LLM student. Hence, the student LLM directly embeds the watermark into its trained weights, with the watermark detector used for distillation being employed to flag watermarked texts. 

Other authors bet on \textit{Reinforcement Learning (RL)}, as \cite{Xu_2024_RL_watermarking} introducing LLM watermarking into the Reinforcement Learning with Human Feedback (RLHF) pipeline of \cite{Ouyang_2022_RLHF_pipeline}. They jointly train the LLM and a watermark detector, optimizing both the quality of LLM and the detectability of the watermark. 

\subsubsection*{Step 2: Weights Editing}\label{itm: In-Processing - Step 2}

Other avenues of watermarking into the model remain open, once the LLM has been trained. The LLM signature can be inserted at the level of LLM parameters, in some or all layers and versions of the model \citep{Bansal_2022_gaussian_noise}. This weights' modification can take the form of adding bias or noise, which is supposed to robustly and discretly impact the future LLM text. 

\cite{zhang2024_quantized_llm_EmMark} focus on the most salient parameters, where the gap between minimal and maximal values on their activation layer reaches the highest magnitude \citep{lin2024_activation_salient_weights}. They add a small bias $\xi$ to these parameters $\theta$ on a random basis: $\theta \to \theta + \xi$. The watermarking stamp displays resistance in the face of attackers, which do not know the bias factor $\xi$ that is also the key. Building on insights from \cite{li2023_weight_quantization}, the authors finally quantize the LLM to make it even harder to attack with the sole int8 version. 

\cite{block2025_gaussmark} also add a slight bias to the parameters $\theta$ of the model, which is a key $\xi$ held by providers. Unlike \cite{zhang2024_quantized_llm_EmMark}, they do not require a quantized version of the LLM. 
Instead of using activation magnitude, the authors empirically choose a single Multi-Layer Perception (MLP) of a Transformer block before the last activation function to preserve the quality of predictions. 
Their bias factor follows a normal distribution centered around 0 and with a small variance $\sigma > 0$, which bear the advantage of negligibly affecting the LLM quality. As Gaussians present rotational invariance and independent directions, the bias signal is shown to be orthogonal to LLM weights: 

$$\xi \sim \mathcal{N}(0,\,\sigma^{2}I)$$

\subsubsection*{Step 3: Next-Token Distribution}\label{itm: In-Processing - Step 3}
Out of the model training and parameters, there are still steps before the model generates its final token $v_t$. The autoregressive model generates a distribution of logits, which are then passed into a softmax, computing the probability each token has to occur in position $(t)$. The prompt $x$ and previous tokens ${v_1, ..,v_{t-1}}$ being provided, the probability of any token to come off $p(.|x\cap v_1,..,v_{t-1})$ varies as shown in Figure \ref{fig:watermarking_during_inference}.

Most watermarking approaches during inference time happen at this moment, a famous one being designed by \cite{Kirchenbauer_green_tokens}. At each generation of token $(t)$, the hash of the preceding token $(t-1)$ seeds a random generator, used to partition the vocabulary $V$ into Red and Green tokens. A bias is then added to the logits of Green tokens, enabling them to be selected more frequently than Red ones for the next-token generation (Cf. Figure \ref{fig:watermarking_during_inference}, Step 3).

On their side, \cite{Aaronson_2022_next_token_sampling} also alter the next-token distribution with a Gumbel-softmax. This Gumbel rule picks up certain stochastically sampled points, which are the only-one whose logits get converted into probabilities. The authors probe their watermarking bias to be \textit{distortion-free}, keeping intact the quality of the final generated content.  

\subsubsection*{Step 4: Next-Token Sampling}\label{itm: In-Processing - Step 4}

From the distribution of tokens' likelihoods to the final token generation, taxonomies often omit the crucial step of next-token sampling. Yet, very different strategies exist to single out $v_{t}$: beam search compares the most probable sequences assembling the token $v_{t}$ with the $(t-1)$ first tokens, while greedy sampling considers a unique sequence and computes a likeliness score for each $v_t \in V$ \citep{Christ_2024_next_token_toy_LLM, Aaronson_2022_next_token_sampling}. Therefore, the selection of the next token $v_t$ can be inflected after the distribution of the candidates has been generated. This is the last step where watermarking might arise during inference time, biasing the sampling or selection procedure. 

In a recent work, \cite{DeepMind_dathathri_2024_complexifying_tokens} combine the biased logits of a specific subset of tokens (Cf. Step 3) with tournament sampling. At each step of the tournament, the authors allocate the tokens of the vocabulary $V$ over $m \in \mathbb{N}$ chosen colors, according to \cite{Kirchenbauer_green_tokens} hashing procedure and the recent context of the prompt $x$. Their algorithm gives each token a pseudo-random score, which is compared to another token's score. The token with the highest score is selected for the next turnament, until the last remaining top-score token is output as $v_t$. 

\newpage
\section{Evaluating Watermarking in the Literature}\label{sec:evaluating_pros_contras}
When comes the time for evaluation, LLM watermarks do not only face up to the diversity of metrics, but also different kind of objectives. For the time being, we put aside the four requirements of the EU AI Act, to be able to better interpret them in Section \ref{sec:links_scient_eu_evaluation} with technical insights. We choose to focus on three pillars which are frequently adopted in LLM literature to gauge watermarks:\footnote{For more details, we turn the reader to \cite{liu2024_text_survey} p. 15, 
 \cite{zhao2025_watermark_overview} pp. 14-15, and \cite{pan2024_markllm_toolkit} pp. 5-6. They depict a wide range of optimization tools and metrics to operationalize the evaluation of LLM watermarking.} 

\subsection{Ensuring Watermark Detectability}\label{sec:criterion_detectability}
Being primarily used to authenticate LLM-generated text, watermarks have to bear some recognizability. To that extent, each LLM watermark we presented in Section \ref{sec:pre_post_processing_watermarking} and \ref{sec:in_processing_watermarking_more_formal} encompasses an integration part, where it is hidden in the model, tokens or text, and a \textit{detection} part. 
Zero-bit watermarks indicate if the text has or not been LLM-generated, when multi-bits incorporate other information on the copyright or generation date. Therefore, the zero-bit approaches we displayed in previous Sections are computationally lighter and represent the most used watermarks \citep{liu2024_text_survey}. For these watermarks, detectors are committed to statistical tests. 

\cite{block2025_gaussmark} implement a test where the null hypothesis corresponds to a negative: the text has not been generated by the LLM put into test. If the null hypothesis is rejected with a sufficiently low p-value, the 5\% confidence interval suggests that the tokens are output by the LLM whose weights have been Gaussian-biased. Closely, \cite{Kirchenbauer_green_tokens} test the null hypothesis that the text has been produced with no knowledge of the separation rule between Red and Green tokens, based on hash function $(t)$. Their claim to reject is that Green tokens in $V$ have not been overly promoted by the iterative hashing process. The authors compute a z-statistic according to this null hypothesis, and state that the content has been generated by the Red / Green watermarked LLM if the z-value exceeds a chosen threshold. 

Beside these evaluations or combined with them, traditional binary classification metrics also serve to detect zero-bits LLM watermarks. For evaluation, they usually compare an equal number of texts written by humans and contents generated by AI. \cite{block2025_gaussmark} priority is to avoid false positives, e.g. to falsely identify someone as a plagiarist in educative or artistic contexts. We have here a specific instance of a dilemma described for years in Signal Detection Theory, between False Positive Rate (FPR) and False Negative Rates (FNR) \cite{mcnicol2005primer}. 
These detectability metrics also reflect moral choices of the one which implement them, depending on the optimal threshold between them determined through Precision-Recall (PR) or Receiver Operating Characteristic (ROC), and on the priority set by deciders between these errors \cite{souverain2024implementing}. They enclose a sense of acceptability, as measurement errors made by the detectors are considered as a price to pay.

In addition to the precision or recall of detection tools, there are practical challenges to their identifiability. Detectors need enough tokens to be confident in asserting that series of words are the fruit of a model. Among next-token distribution techniques (step 3), \cite{KGW_minimal_length} exhibit \cite{Kirchenbauer_green_tokens} to require the minimal length of tokens for stable watermark distribution, at a FPR of 2\%. 
A related and tough issue is to enable detection after subtle modifications of the LLM-text, overlooked by the reviews we inspected on LLM watermarking. The concrete example of shwred cheaters must urgently be considered in education, with small insertions, replacements, deletions or copy-pastes. 
When does inspiration move to plagiarism, and vice versa? With objective clues as the level of alteration of an initial text, interpretations of detection measures should be aware of a subjective one, linked with the personal intention, honesty and use of the human embedding LLM into one’s content. 

\subsection{Ensuring Watermark Robustness}\label{sec:criterion_robustness}

Instead of the detectability of watermarks, other authors stress on their \textit{undetectability}. Without the private key, attackers are not supposed to recognize a watermarked content. This change of perspective is relevant in contexts where the model's owner highlights cybersecurity, copyright and protection of one's intellectual property \citep{Meta_taxonomy_fernandez, Christ_2024_next_token_toy_LLM}. Maleficent actors claiming they used a specific LLM to produce wrong knowledge or reprehensible beliefs, shall not be able to mark their content with this LLM authentication stamp. 

Hence, we see as a second major requirement for watermarks to be robust or resilient to changes. These changes include attacks which are \textit{targeted} against specific LLMs, such as spoofing which may imitate the Red / Green coloring of tokens of \cite{Kirchenbauer_green_tokens}. Attackers might also append minimal changes in words, unicodes or tokens, in order to use an LLM-content without being flagged. This kind of attack is not watermark-specific, or \textit{untargeted} \citep{liu2024_text_survey}.

These two risks stand at the exact symmetric, focusing either on taking possession of the watermark or the LLM (see our schema in Table \ref{tab:schema_watermark_stealings}). On one side, spiteful agents mimick a targeted LLM watermark to \textit{wrongly attribute} an illegitimate output to this. They might discard the developer while spreading fraudulent or antisemitic contents allegedly generated by its LLM. The first side \textit{steals the watermark}. On the other side, plagiarists \textit{tracelessly steal} the LLM output. They benefit from the windfall effect of fast and accessible generated text, as described above in education, while erasing any watermarking sign. The second side \textit{steals the LLM output}.
While the first resulting defense stresses on protecting the LLM author and integrity \citep{zhao2025_watermark_overview}, that must not be imitated as shown in Table \ref{tab:schema_watermark_stealings}, the second one aims at preserving its signature that shall not be easily erased. 


\begin{table}[!ht]
    \centering
    \fontsize{10pt}{10pt}\selectfont

    \caption{Defending the LLM: Two Criteria to make the Watermark Robust}
    \label{tab:schema_watermark_stealings}
    \renewcommand{\arraystretch}{1.6} 
    \setlength{\extrarowheight}{3pt}  

    \setlength{\tabcolsep}{15pt} 
    \begin{tabular}{|>{\raggedright\arraybackslash}p{4.2cm}|
                      >{\raggedright\arraybackslash}p{4.2cm}|
                      >{\raggedright\arraybackslash}p{4.2cm}|}
\hline
    \rowcolor{gray!25}
    \multicolumn{1}{|c|}{\textbf{Watermark Robustness Criteria}} &
    \multicolumn{1}{c|}{\textbf{Target of the Attacker}} &
    \multicolumn{1}{c|}{\textbf{Malicious Attacking Use}} \\ \hline

    \textbf{Non-Extractable} &
    Watermarking method &
    Attackers can insert their content, pretending to be generated by the defamed LLM \\ \hline

    \textbf{Non-Erasable} &
    LLM-generated text &
    Attackers can insert LLM-generated text, pretending to be their own \\ \hline
    \end{tabular}
\end{table}

In the spirit of attacks seeking to steal the LLM while wiping out the watermark, 
the resilience of watermarks must be proved against operations which take place during the LLM lifecycle.  \cite{gloaguen2025_open_source} and \cite{Wu_2023_taxonomy_resiliency} advocate for LLMs to remain detectable once quantization, pruning, weights' merging and fine-tuning has been applied to the originally watermarked LLM. This injunction is crucial in open-source contexts: once the model has been trained and watermarked, further alterations of its weights or architecture shall keep the watermarking signal trackable.\footnote{We see these further operations in open-source contexts as equivalent to fine-tuning or distillation attacks described in \cite{Wu_2023_taxonomy_resiliency, liu2024_text_survey} Being described as attacks or natural LLM steps, they follow the same direction: preserving the watermarking signal despite further steps of LLM development. }
The watermark detector must remain functional, after intentional or non-intentional modifications on LLM structure and outputs. 

\subsection{Maintaining LLM Quality}\label{sec:criterion_quality}

Even if the LLM has been shown to be detectable and robust against external changes, it won't be used if the watermark heavily affects the quality of LLM-generated contents. 

In that respect, \textit{comparative} tools exist to measure the difference of quality between unwatermarked and watermarked outputs for the same LLM. Meteor or BLEU Scores primarily used in translations, are usually deployed to penalize the watermarked LLMs when their answers to prompts differ in precision or length from the non-watermarked output \citep{BLEU_evaluation}. As such metrics are sensitive to word order and the initial structure of the text, they can be completed with \textit{semantic} evaluation tools. For instance, \cite{Yoo_2023_inference_quality} have implemented Sentence-BERT \citep{Sentence_BERT_semantic_quality} to compute cosine similarity between semantic embeddings. Introducing an Entailment Score, the authors try to further capture if logical relationships are preserved in the watermarked answer. 

Single-text evaluation metrics also exist to assess the quality of watermarked texts. They do not merely differ from the classic quality evaluation of trained LLMs, using perplexity or human scoring \citep{Yoo_2023_inference_quality}. On isolated tasks, the watermarked LLMs are evaluated with standard evaluation methods for LLMs: e.g. BLEU and BERTScore for translation and text summarization \citep{BLEU_evaluation}, perplexity or logarithmic diversity for text completion, GPT-Truth and GPT-Info for question-answering and fact-checking \citep{Chen_2023_GPT_Truth}, or LLM-as-a-judge for instruction following \citep{Zhao_2023_watermarking_LLM_as_a_judge}. 

While these metrics classically evaluate the quality of LLM ouputs (i) compared to the non-watermarked LLMs and (ii) for specific tasks, some authors also put emphasis on (iii) diversity of the generated answers. In-Processing watermarking methods in Step 3 and Step 4 rely on biased selection of some tokens (Cf. Section \ref{itm: In-Processing - Step 3}), which can reduce the lexical and semantic diversity of LLM texts. \cite{Kirchenbauer_green_tokens} themselves note this default of their token-coloring approach. They propose both a solution, implementing Red / Green tokening only for high-entropy tokens, to introduce distortion for words where the diversity of synonym expressions will not compromise the text quality, and a logarithmic measure of the diversity produced by an LLM \citep{Kirchenbauer_criticizing}. 

Other metrics evaluate the loss of diversity produced by watermarking, such as n-gram repetition of tokens. With this gauge, \cite{Gu_2023_distillation_watermarking} compared two watermarks happening at the same step of LLM construction, through biasing the logits' distributions. (see Step 3 of Section \ref{itm: In-Processing - Step 3}). They found that \cite{Kirchenbauer_green_tokens} method was of less impact for the content's diversity, compared with the Gumbel approach of \cite{Aaronson_2022_next_token_sampling}. 


\section{Watermarking in Europe: Operational Interpretation of the AI Act}\label{sec:links_scient_eu_evaluation}

Now we presented the technical ways LLM watermarks are currently evaluated, how can we relate them with the four criteria defined in the EU AI Act\footnote{Cf. AI Act \cite{AI_Act_Europe}, Article 50, Recital 133, and our Introduction \ref{sec:intro_eu_ethical_and_market_needs} for more context.}: reliability, interoperability, effectiveness and robustness?

\begin{figure}[h!]
\caption{Interpreting the AI Act Criteria for LLM Watermarking: From Overlaps to Measurable Standards}
\label{fig:overlap_eu_criteria}
\centering
\includegraphics[scale=0.7]{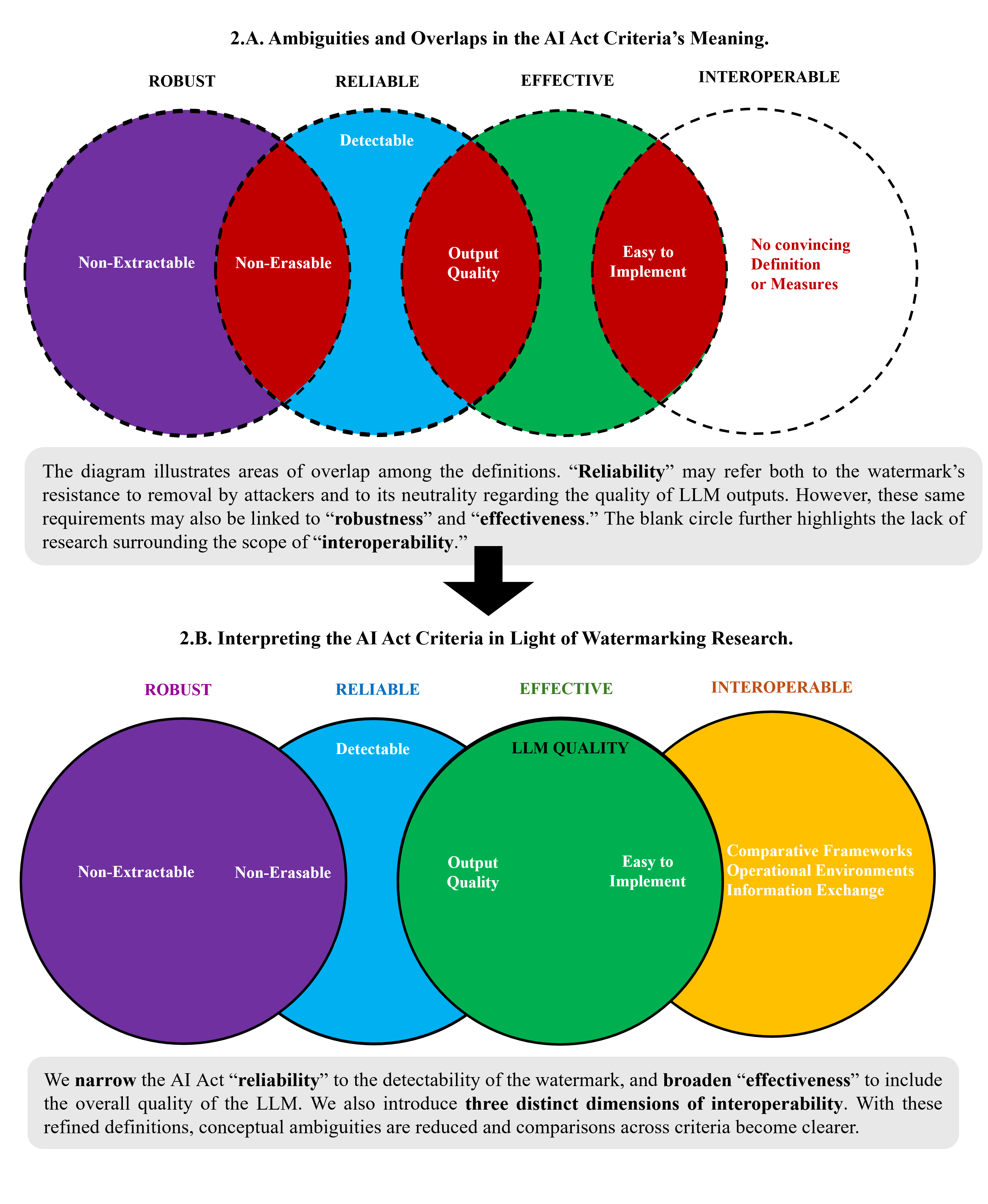}
\end{figure}

Depending on how we define these four EU claims, one may observe overlaps between them. This is a necessary consequence of their elusive mention in Recital 133, which is also not made explicit in the recent transparency chapter of the Code of Practice for the providers "General-Purpose AI" - including LLMs \citep{Code_of_Practice_GPAI_Europe}. We pictured the overlaps in Figure \ref{fig:overlap_eu_criteria}.A, which may lead LLM providers and evaluators to unpleasant confusion when assessing AI conformity to the AI Act. To fill the gap, the following Section provides \textit{clear and operational interpretations} of the EU evaluation pillars with state-of-the-art, reachable, and distinct objectives, outlined in Figure \ref{fig:overlap_eu_criteria}.B. Our refinement links the AI Act with the context of LLM watermarks, enabling clear guidance of LLM providers. 

\newpage
\subsection{Interpreting "Reliability" as Detectability}\label{sec:eu_keep_robustness}

"Robustness" is the only European criterion we find as such in watermarking literature, if the AI Act denotes robustness of the watermark \textit{under attacks} - as it is in most evaluation research \citep{liu2024_text_survey,pan2024_markllm_toolkit}. It is comprised of the two symmetric attacks we analyzed in Section \ref{sec:criterion_robustness}, Table \ref{tab:schema_watermark_stealings}. The watermark must not be \textit{extracted} by malicious agents wanting e.g. to attribute shameful contents to institutional chatbots ; it must not be \textit{erased} by dishonest people, claiming they wrote a piece of text that is in fact fully synthetic.

Lappings arise with "reliability". What is, indeed, reliable: the watermarking method, or the LLM that embeds the watermark? The Recital 133, brief on the matter, does not elucidate it. Let us consider that reliability applies primarily to the watermark technique, which makes sense in the enumeration of the AI Act: \textit{the technique} shall be robust, the technique shall be reliable, effective... 
An obvious sense is that the technique can be \textit{reliably detected}, reaching sufficient confidence intervals, p-values or z-scores indicated in Section \ref{sec:criterion_detectability}. 

However, one could also mean that the watermark is reliable if its signal passes through possible transformations along the LLM lifecycle - training, deployment, fine-tuning, Retrieval Augmented Generation (RAG) until the final generation of tokens \citep{gloaguen2025_open_source}. This reliability along LLM changes \textit{overlaps with a robustness dimension}, which is, the watermark signal must not be erased.\footnote{See Figure \ref{fig:overlap_eu_criteria}.A, left red intersection.} 

Therefore, we suggest to avoid confusion in the EU criteria and \textit{to interpret "reliability" as detectability of the watermark}. Using the most restrictive sense as possible to interpret "reliability", that is focusing only on statistical detectability of the signal (tests on various architectures, lengths and combinations of tokens, relevance of the corpus...) constitutes a piece of work which deserves its own category.\footnote{See the clarified concept of detectability in Figure \ref{fig:overlap_eu_criteria}.B, blue.}

\subsection{Interpreting "Effectiveness" as LLM quality}\label{sec:eu_effectiveness}

The overlaps continue with “effectiveness".  A technique is considered effective when it achieves the goals intended by its designers. It \textit{effectively} does what it was programmed for. As we observed in Section \ref{sec:evaluating_pros_contras}, these goals are threefold: robustness, detectability, and LLM quality. To maintain distinct AI Act criteria, we already treated robustness and detectability as separate categories. What remains is LLM quality as the defining aspect of effectiveness.

As noted in Section \ref{sec:eu_keep_robustness}, reliability was ambiguous: it could refer either to trustworthy watermarking or to faithful model outputs.\footnote{See Figure \ref{fig:overlap_eu_criteria}.A, red, right overlap between reliability and effectiveness.} Defining LLM quality as an independent criterion helps resolve this ambiguity between “reliability” and “effectiveness”. Reliability now evaluates the \textit{tool}, while quality concerns the \textit{outputs}.

Finally, if effectiveness is understood as keeping the LLM relevant for users, quality must also account for implementation factors. Beyond the perceived quality of outputs, ease and cost of implementation are relevant. This extended reading of “effective” watermarking thus combines qualitative outputs and implementable LLM under the unified notion of LLM quality.
\footnote{See the clarified concept of LLM quality in Figure \ref{fig:overlap_eu_criteria}.B, green.}

\newpage
\section{Interpreting "Interoperability": Three Research Directions}\label{sec:eu_interoperability}

Interoperability warrants a separate discussion, as its technical foundations remain unclear and largely unexplored in current LLM research. Literally, watermarks are interoperable if they can \textit{work together} to authenticate AI-generated contents.\footnote{Cf. Cambridge English Dictionary, ed. 2025, consulted on 2025/10/27, \hyperlink{https://dictionary.cambridge.org/dictionary/english/interoperable}{"Interoperable"}.}
We highlighted above three working directions in which marking techniques have then to converge: keeping the signal robust, and detectable, without deteriorating the LLM quality. Besides, operability points towards operationalization: watermarks must also be easily embedded in different forms of LLMs, and comparable with other authentication methods. 

We could imagine a global detector for all LLMs registered on the European market with their compulsory watermarks, made accessible to public deciders. Such \textit{objective} instruments would help EU auditors to check for LLM compliance. This ideal global platform would also complement the \textit{subjective} intuition of teachers. Interoperable watermarks would be detectable regardless of the marking style, the original provider (Google, OpenAI, Anthropic...), and the potential transformations of an LLM (RAG, fine-tuning, mixed architectures...). For a teacher, it would  indicate above reasonable thresholds that the dissertation of any suspected student was LLM-generated. 

However, except emerging comparisons of LLM watermarks narrowed to \textit{LLM quality} across different models \citep{dang2025_inter_models_perf_evaluation, gloaguen2025_open_source}, existing literature leaves an unexplored area within the field of interoperability. The concept is not yet installed in LLM watermarking comparison. To bridge the gap, we sketch below three dimensions in which LLM research should urgently work to give birth to global evaluations on interoperability:

\subsection{Comparative Frameworks}
Evaluating \textit{every technique}, including fingerprints and logging mentioned in the same Recital 133 \citep{AI_Act_Europe}, must be covered by interoperability. To combine them adequately, we have to compare all the methods authenticating LLM contents across robustness and detectability of the signal, and LLM quality. 
    Open-source toolkits already exist to compare \textit{some styles of watermarks} within each other on these three main requirements. For instance, MarkLLM \citep{pan2024_markllm_toolkit} regularly actualizes a GitHub repository to implement and compare Next-Token Distribution based on the coloring of tokens \citep{Kirchenbauer_green_tokens} (see Section \ref{itm: In-Processing - Step 2}) and Next-Token Sampling inspired from cryptography \citep{Christ_2024_next_token_toy_LLM} (see Section \ref{itm: In-Processing - Step 4}). Whereas the MarkLLM benchmark compares methods from different styles and with replicable code, more general initiatives shall be led to evaluate marking techniques in a complete manner. 

\subsection{Operational Environments} 
LLM Watermarking techniques do not only have to prove their detectability, robustness and least impact on LLM quality compared with other authentication tools. They must also achieve the three goals \textit{once LLM are deployed}, and used on a daily basis. In this perspective, the method of DeepMind, inspired from \cite{Kirchenbauer_green_tokens} coloring to pseudo-randomly sample the tokens (see Section \ref{itm: In-Processing - Step 4}, \cite{DeepMind_dathathri_2024_complexifying_tokens}), was tested with 20 million Gemini responses. We do not however consider it as sufficient to prove that a LLM watermark is ready for large-scale production. The authors should also extend their experiments beyond a fixed 5\% False Positive Rate \citep{block2025_gaussmark}, testing detectability and robustness under conditions such as quantisation, pruning, merging, and fine-tuning, common in open-source LLMs \citep{gloaguen2025_open_source}. Finally, the method has recently shown to be vulnerable to textual attacks on semantics, as paraphrasing and back-translation \citep{han2025_google_synthid_attacks_semantic_text}. Even with this rare example of large-scale testing in LLM watermarking, we call for more comprehensive and realistic frameworks that support integration at scale and in real-world digital environments.

\subsection{Information Exchange} 
Lastly, interoperability may compare authentication methods \textit{in the way they share information}. Just as detectives using different methods share clues, fingerprinting and watermarking techniques may prove complementary in ensuring the authenticity of LLM outputs. Taking this further, the Coalition for Content Provenance and Authenticity (C2PA), founded in 2021 by a consortium of technological companies, introduced an international specification which lists robust ways of signing a document and attributing its content for these industrials, \textit{mainly coming from cryptography} \citep{C2PA_specification}. 
Even if these "hard binding" metadata prove the authenticity of the content, often with a unique hashing signature, metadata might be lost after compressions or conversions, and complex to recover. That is why the C2PA conceptualized "soft binding", like \textit{watermark} or \textit{fingerprint}, whose embedded and invisible mark would help to ensure the authenticity of LLM contents.\footnote{See \cite{C2PA_specification}, 9.3, "Soft Bindings".} C2PA pioneer standards stress that marking methods mentioned by the EU AI Act \citep{AI_Act_Europe} \textit{are} complementary. Further research should explore \textit{how} these methods can work together to improve output attribution - for instance, by jointly narrowing the confidence interval.

\subsection{Foundations for future works on Interoperability}
Probably because LLM watermarking techniques mostly lie at the stage of advanced research, interoperability suffers from the absence of discussions on industrialized articulation of techniques, measurements, and even definitions. In that sense, the three dimensions we suggested above mark a progress to clarify the EU AI Act requirement of interoperability.\footnote{See the Figure \ref{fig:overlap_eu_criteria}.B, with the yellow circle of "interoperable" LLM watermarks filled with our three dimensions of global, at-scale, and information-sharing comparison.}

The field has to stay tuned with rapidly evolving LLMs and associated environments, and implementation aspects will certainly lead to more solid operational assessments in future months and years. 
However, experimental measurements of the three dimensions of interoperability remain rare and only partially explored: they fall outside the scope of current LLM watermarking benchmarks \citep{Kirchenbauer_criticizing, liu2024_text_survey,zhao2025_watermark_overview}. Hence, interoperability cannot yet be integrated into the evaluation presented in Section \ref{sec:overall_strengths_weaknesses} and Table \ref{tab:strengths_weaknesses_watermarks_llm_overview}. Doing so would not be supported by sufficient empirical evidence.

\section{Trade-Offs for Existing LLM Watermarking Techniques}\label{sec:overall_strengths_weaknesses}


We interpreted the EU criteria in a rigorous and measurable way. How can we connect them with the state-of-the-art techniques and evaluations introduced above? This last section is dedicated to comparing the LLM watermarking methods along the EU evaluation axes. As the final Table \ref{tab:strengths_weaknesses_watermarks_llm_overview} summarizes, each style of watermarking bears its own advantages and defaults. In particular, every watermarking family presents to LLM providers its own trade-offs between robustness, detectability, and effectiveness. 


\newcommand{\outerrule}{0.8pt}  
\newcommand{\innerrule}{0.5pt}  

\begin{sidewaystable}[p]
\fontsize{12pt}{12pt}\selectfont
\centering
\caption{\textbf{The strengths and weaknesses of LLM watermarking styles.}
In green cells are criteria met with high confidence, on the contrary to red.
Yellow cells suggest a medium maturity of the technique.\\
}
\vspace{-2mm}

\renewcommand{\arraystretch}{1.6}
\setlength{\tabcolsep}{10pt}


\resizebox{\linewidth}{!}{%
\begin{tabular}{!{\color{black}\vrule width \outerrule}
                m{4cm}
                !{\color{gray!50}\vrule width \innerrule}
                m{2.2cm}
                !{\color{gray!50}\vrule width \innerrule}
                m{2.2cm}
                !{\color{gray!50}\vrule width \innerrule}
                m{2.9cm} 
                !{\color{gray!50}\vrule width \innerrule}
                m{2.4cm}
                !{\color{gray!50}\vrule width \innerrule}
                >{\columncolor{white}}m{2.4cm}
                !{\color{black}\vrule width \outerrule}}
\arrayrulecolor{black}\hline

\multicolumn{1}{|>{\centering\arraybackslash}m{4cm}|}{\textbf{Evaluation Criteria ******************* Point of LLM life-cycle where Watermark is developed}} &
\multicolumn{2}{>{\centering\arraybackslash}m{5cm}|}{\makecell{\textbf{ROBUST}\\\textbf{Watermark}}} &
\multicolumn{1}{>{\centering\arraybackslash}m{2.9cm}|}
{\makecell{\textbf{DETECTABLE}\\\textbf{Watermark}}} &
\multicolumn{2}{>{\centering\arraybackslash}m{5.5cm}|}{\makecell{\textbf{EFFECTIVE}\\\textbf{LLM}}} \\
\hline

\multicolumn{1}{|>{\centering\arraybackslash}m{4cm}|}{} &
\multicolumn{1}{>{\centering\arraybackslash}m{2.2cm}|}{\textbf{Non-Erasable}} &
\multicolumn{1}{>{\centering\arraybackslash}m{2.2cm}|}{\textbf{Non-Extractable}} &
\multicolumn{1}{>{\centering\arraybackslash}m{2.9cm}|}
{\textbf{Detectable}} &
\multicolumn{1}{>{\centering\arraybackslash}m{2.4cm}|}{\textbf{Easy to Implement}} &
\multicolumn{1}{>{\centering\arraybackslash}m{2.4cm}|}{\textbf{Output Quality}} \\
\hline


\multicolumn{6}{!{\color{black}\vrule width \outerrule}l!{\color{black}\vrule width \outerrule}}{\textbf{Pre- \& Post-processing}} \\
\hline
\rowcolor{gray!5} Direct replacement of words or characters
& \cellcolor{red}   & \cellcolor{red}   & \cellcolor{green}
& \cellcolor{green} & \cellcolor{green} \\
\hline
\rowcolor{gray!15} Selective replacement (random, contextual...)
& \cellcolor{green} & \cellcolor{red}   & \cellcolor{green}
& \cellcolor{red}   & \cellcolor{green} \\
\hline

\multicolumn{6}{!{\color{black}\vrule width \outerrule}l!{\color{black}\vrule width \outerrule}}{\textbf{Next-Token Distribution}} \\
\hline
\rowcolor{gray!5} Small Window Size
& \cellcolor{green} & \cellcolor{red}   & \cellcolor{green}
& \cellcolor{green} & \cellcolor{red}   \\
\hline
\rowcolor{gray!15} Large Window Size
& \cellcolor{red}   & \cellcolor{green} & \cellcolor{yellow}
& \cellcolor{yellow}& \cellcolor{green} \\
\hline

\multicolumn{6}{!{\color{black}\vrule width \outerrule}l!{\color{black}\vrule width \outerrule}}{\textbf{Next-Token Sampling}} \\
\hline
\rowcolor{gray!5} Adding bias to specific blocks of tokens
& \cellcolor{yellow}& \cellcolor{red}   & \cellcolor{green}
& \cellcolor{red}   & \cellcolor{green} \\
\hline
\rowcolor{gray!15} Mapping random numbers to LLM samples
& \cellcolor{green} & \cellcolor{yellow}& \cellcolor{yellow}
& \cellcolor{yellow}& \cellcolor{yellow}\\
\hline

\multicolumn{6}{!{\color{black}\vrule width \outerrule}l!{\color{black}\vrule width \outerrule}}{\textbf{Into LLM Architecture}} \\
\hline
\rowcolor{gray!5} Distillation of a watermarked teacher model
& \cellcolor{green} & \cellcolor{red}   & \cellcolor{green}
& \cellcolor{yellow}& \cellcolor{green} \\
\hline
\rowcolor{gray!15} RL with a watermark detector
& \cellcolor{green} & \cellcolor{yellow}& \cellcolor{yellow}
& \cellcolor{yellow}& \cellcolor{green} \\
\hline
\rowcolor{gray!5} Adding Bias on specific LLM weights
& \cellcolor{green} & \cellcolor{yellow}& \cellcolor{green}
& \cellcolor{green} & \cellcolor{green} \\
\arrayrulecolor{black}\hline
\end{tabular}
}

\label{tab:strengths_weaknesses_watermarks_llm_overview}
\end{sidewaystable}

\subsection{Pre- and Post-Processing: slow detection or easy removal}

In Section \ref{sec:pre_post_processing_watermarking}, we mentioned a set of methods interested in training data and embeddings (pre-processing) or already generated text (post-processing). These techniques enforce subtle additions, deletions, or substitutions into textual elements, being or not vectorized. These schemes do not meddle with the inner model, in contrast to in-processing methods which involve specific training \citep{Xu_2024_RL_watermarking}, modifications in tokens \citep{Aaronson_2022_next_token_sampling, Kirchenbauer_green_tokens} or architectural changes into LLM layers \citep{Bansal_2022_gaussian_noise, block2025_gaussmark}. To the extent they do not affect the generation process, most pre- and post-processing approaches preserve the \textit{global quality} of the unwatermarked LLM outputs \citep{liang2024watermarking}. 
Besides, introducing hidden characters \citep{Wei_2024_pre_character_substitution_pseudo_random,Sato_2023_post_character_subst_EASYMARK} or synonyms \citep{Zhang_2024_post_nn_syntax_replacement} 
allows for understandable substitution and straightforward detection. 

In spite of that, clear-cut characters' and synonyms' insertions are vulnerable to simple attacks like random synonym substitution. Incursive actors may simply query LLM outputs through Application Programming Interfaces (APIs), replacing or erasing the words and Unicodes put forward during watermarking \citep{liu2024_text_survey}. 
Watermarking can be spoofed in pre-processing methods \citep{Wei_2024_pre_character_substitution_pseudo_random}, where the signal embedded in the training data cannot be changed without retraining the LLM.

Yet, synonym substitutions which are pseudo-random \citep{Wei_2024_pre_character_substitution_pseudo_random} or sensitive to the context \citep{Zhang_2024_post_nn_syntax_replacement} show more resistance to textual attacks. This is also the case for methods entwining syntactic and semantic modifications, such as \cite{Abdelnabi_2021_post_decoder_adversarial_replacement} using adversarial training to introduce substitutions, which become especially hard to be disclosed by attackers. 

In all of these cases, however, the \textit{robustness to textual attacks} arises at the expense of \textit{heavier implementation}. Involving a rejection sampling algorithm to position synonyms into $k$ semantic spaces, $k-$SEMSTAMP winds down the generation time of the text. The detectors in \cite{Zhang_2024_post_nn_syntax_replacement} and \cite{Abdelnabi_2021_post_decoder_adversarial_replacement} need a specific classification training, which is also computationally costly. As the resistance to learnability (see Section \ref{sec:criterion_robustness}) of such watermarks shall be studied over simple textual attacks, this first, superficial layer of watermarking robustness compromises the quality of LLMs. 

\subsection{In-Processing Approaches}
In contrast, in-processing methods mark LLMs with a non-superficial stamp, as they embed their watermark during the generation time of LLMs. 
As shown in this section and summarised in the last three categories of Table \ref{tab:strengths_weaknesses_watermarks_llm_overview}, in-processing watermarking strikes a balance between strong resistance to the signature being extracted or erased and increased computational load on the LLM. This additional burden can sometimes lead to a loss in output quality.

\subsubsection*{Next-Token Distribution (step 3): Extractable or Erasable}\label{sec:next_token_distrib_strength}

In that manner, printing the next-token distribution lies in the foreground of LLM watermarking. 
Happening after the training of the LLM, such approaches do not alter the inner architecture but the logits output by the model. They are detectable to the extent they add a slight bias to certain tokens, which is also the model's signature $\xi$. Intuitively, a trade-off arises: the \textit{detectability} of $\xi$ among tokens might hit resistance to imitation or plagiarism attacks (less \textit{robustness}). 


A key setting is the window size,\footnote{See Table \ref{tab:strengths_weaknesses_watermarks_llm_overview} for an picture of the advantages borne by large or small window size in Next-Token Distribution watermarking.} i.e. the number of preceding tokens which are used to generate the selection of biased tokens at each generation step $(t)$. A large window size makes \cite{Kirchenbauer_green_tokens} and \cite{Aaronson_2022_next_token_sampling} signatures more difficult to steal, as the dependencies between tokens become more complex. However, any targeted textual attack as rewording gets more chances to break these dependencies, erasing the watermark. The reverse is true for small window sizes \citep{liu2024_text_survey}: the same distribution-based techniques become easier to retrieve or learn from a sufficient amount of watermark-generated text \citep{Gu_2023_distillation_watermarking}, but they gain resistance to paraphrasing as their coloring and separation rules depend on fewer preceding tokens. 

Our third criterion of LLM quality joins the trade-off, as larger window sizes enhance the diversity of generated texts \citep{zhao2025_watermark_overview}. \cite{Kirchenbauer_green_tokens} focus on soft tokens, with high entropy, is also shown to be less compromising for the diversity and human satisfaction on LLM contents, at the cost of more complicated procedures to detect the watermark \citep{Kirchenbauer_criticizing}. 

\newpage
\subsubsection*{Next-Token Sampling (step 4): Balancing LLM Quality and Undetectability} \label{sec:next_token_sampling_strength}

Like logits' techniques, next-token sampling watermarking introduces a bias into the generation process. It guides the selection of the next token $v_t$, associating pseudo-random numbers to privilege some tokens \citep{Christ_2024_next_token_toy_LLM,Aaronson_2022_next_token_sampling, Kuditipudi_2023_sampling_pseudo_random_KTH} or semantic embedding spaces \citep{Hou_2024_k_SemStamp}. 


Though understudied, sampling watermarking shows divergent strengths and weaknesses depending on the methods. Inspired from cryptography, \cite{Christ_2024_next_token_toy_LLM} progressively sample tokens in one group and, once the block gains enough entropy, associate a unique hashing signature to the block. This approach stresses on \textit{undetectability} or robustness to unallowed imitation, while maintaining a high \textit{diversity} of produced contents on the quality side. However, this two-bits sampling process is computationally heavy to implement in real LLMs and less robust to spoofing or textual replacement attacks. k-SemStamp \citep{Hou_2024_k_SemStamp} whose biased sampling relies on similar semantic spaces, embedding regions of tokens through k-means, presents analogous advantages and defaults. While this method ensures the quality of LLM answers, it is also costly to implement and sensitive to textual attacks in targeted semantic regions. 

On the contrary, \cite{Aaronson_2022_next_token_sampling} and \cite{Kuditipudi_2023_sampling_pseudo_random_KTH} samplings can be deployed in LLMs out of toy-examples. Against removal attacks, \cite{Kuditipudi_2023_sampling_pseudo_random_KTH} proposed to enhance the token-sampling of \cite{Aaronson_2022_next_token_sampling} by extending the pseudo-random number sequence over the text length. However, both of these token-based sampling might easily be learned by an attacker model from watermarked-generated data, especially in low-entropy scenarios
\citep{Gu_2023_distillation_watermarking}. Additionnally, if they overly boost some tokens or words, they can compromise the text's \textit{quality} and \textit{diversity}.


\subsubsection*{Watermarking in Model Architecture (steps 1 \& 2): Towards Extractability in open-LLM contexts}


Research on LLM watermarking has so far focused mainly on token-level and text-level methods, whether syntactic or semantic. Approaches that directly modify LLM weights or architectures remain relatively rare. Though, they represent a highly promising direction, that needs to be explored through more development and testing:

\textbf{High quality, embedded detectability and resistance to most targeted attacks.}

So far, the advantages and limitations of such structural approaches have received little systematic attention. For instance, distillation is absent from the survey by \cite{liang2024watermarking}, and mentioned only as an attack strategy in \cite{liu2024_text_survey}. Nonetheless, these methods exhibit several noteworthy advantages. Most of these in-processing techniques integrate the watermark \textit{into the LLM parameters}. In \cite{Xu_2024_RL_watermarking}, the watermark detector is a reward model jointly trained with the LLM. Therefore, the objective of detectability is directly integrated in the trained LLM weights. \cite{Bansal_2022_gaussian_noise} or \cite{block2025_gaussmark} bias $\xi$ is added to the parameters $\theta$ after training, but this noise modifies only some Transformer blocks, layers, and non-linear functions. Hence, both of these watermarkings into weights do affect the quality of predictions in a reduced manner, compared with data modification (pre- and post-processing) and the biasing of next-token selection (see Section \ref{itm: In-Processing - Step 3}, Steps 3 and 4 of In-Processing Watermarking). 

In contrast, the third approach we mentioned into LLM architecture (Step 1), watermark distillation, is more prone to degrading prediction quality. In \cite{Gu_2023_distillation_watermarking}, the teacher model is a LLM that integrates the watermark either during next-token distribution \citep{Kirchenbauer_green_tokens, Aaronson_2022_next_token_sampling} or during sampling \citep{Kuditipudi_2023_sampling_pseudo_random_KTH}. The decoder that detects the over-coloured or over-sampled tokens is then directly learnt by the student LLM. Thus, the student model inherits from the potential decrease in quality of the watermarked parents, depending on their window sizes (Section \ref{sec:next_token_distrib_strength}) or repetitive sampling (Section \ref{sec:next_token_sampling_strength}, \cite{Gu_2023_distillation_watermarking}). 

At the same time, all three approaches integrate their watermarks into the \textit{structure} of models, which makes them particularly hard to remove from text or tokens. Distilled watermarks of \cite{Kirchenbauer_green_tokens} or \cite{Kuditipudi_2023_sampling_pseudo_random_KTH} are more resistant to textual attacks as paraphrasing than the initial token-watermarking methods \citep{Gu_2023_distillation_watermarking}. Other watermark techniques into LLM architecture also offer to be accurately detectable, thanks to the trained detector model \citep{Xu_2024_RL_watermarking} or a simple back-propagation into the layer where the Gaussian weights are stored \citep{block2025_gaussmark}, provided the key $\xi$ and prompt $x$ reject the null hypothesis (see Section \ref{sec:criterion_detectability}). 

\textbf{Research avenues on final detectability and distortion of LLM outputs.}

Though these structural watermarkings are supposed to make the watermark's detection easier, as the LLMs are specifically trained or modified for that purpose, their ultimate detectability is not guaranteed. The main challenge is to remain detectable \textit{after further LLM modifications}. 

These modifications might have different origins, occurring during the model's development or being due to spiteful acts (see Section \ref{sec:criterion_robustness}). \cite{gloaguen2025_open_source} rightly point that in open-source contexts, the trained LLMs are subject to further refitting. The authors lead a battery of tests, and confirm the overall resistance of in-model watermarking methods to pruning and quantization. Their resistance fits with the role of pruning and quantization, designed to lighten the original model without changing the quality of its completions. 

However, distilled token-distribution and token-sampling watermarkings become harder to distinguish after parameters' merging and fine-tuning \citep{gloaguen2025_open_source}.
\cite{block2025_gaussmark} and \cite{Xu_2024_RL_watermarking} watermarks are also sensitive to merging and fine-tuning modifications, except when fine-tuning happens on specific data as mathematical instructions.  

We deplore that \cite{gloaguen2025_open_source} only measure the detectability through two metrics, namely perplexity and TPR \textit{for a fixed FPR of 5\%}. Together with the level of \textit{distortion} brought in final predictions by in-processing biases \citep{block2025_gaussmark}, these tests should be extended for various ranges of FPR, and for broader detectability criteria (see Section \ref{sec:criterion_detectability} concerning the essentials). The deepness of such in-processing watermarks, into weights and layers, leads to an open issue. 
If the watermarking is too deeply entangled with the LLM architecture, will it be realistic to reliably detect it in the final LLM text? 

We call for further research on this promising avenue, which bears the advantage of low quality impact on text quality and difficult removal of the signal at the weights' level. Tests on detectability and broader integration on computing infrastructures, meant in interoperability, will help to verify if embedding watermarks directly into LLM architecture ensures to reach the three pillars.

\section{Conclusion: Limitations and Further Directions}
Having anchored the EU AI Act requirements in concrete watermarking techniques for LLMs, we proposed an interpretation that is both clear and technically accurate. We refined the four pillars of robustness, reliability, effectiveness, and interoperability, legally requested in Article 50 and Recital 133 for generative AI marking. Thanks to a precise interpretation of "reliability", through the lens of detectability of the \textit{watermark}, and a broad interpretation of "effectiveness", regarding the impact of the watermark on the \textit{LLM}, we propose operational standards to avoid confusion and support guidance of LLM providers. 

An accessible evaluation of watermarking techniques as in Table \ref{tab:strengths_weaknesses_watermarks_llm_overview} is crucial to enable EU governance. Yet, we encourage to develop this table in two directions:

\begin{itemize}
    \item 
    We purposed to give a panorama of watermarking \textit{styles}. To that end, we grouped techniques that are heterogeneous on some points. For instance, inside methods which both use pseudo-random sampling (step 4), \cite{Aaronson_2022_next_token_sampling} hurts more the text diversity than \cite{DeepMind_dathathri_2024_complexifying_tokens}, whose hashing procedure turns out to be more sensitive to paraphrase attacks \citep{han2025_google_synthid_attacks_semantic_text}. To enable this kind of nuances to appear, a solution would be to apply the same categories than in Table \ref{tab:strengths_weaknesses_watermarks_llm_overview}, 
not only for watermarking styles, but also for pairwise comparisons of watermarks. 

\item 
The fourth AI Act pillar of interoperability must be evaluated in our comparative table. However, due to the lack of empirical evidence, we were not able to include interoperability evaluations of watermarking families. LLM research must urgently provide complete comparisons of all marking methods in realistic environments, towards the directions we pointed in Section \ref{sec:eu_interoperability}. Assessing interoperability will help to confirm the robustness of techniques emerging inside LLM architecture, involving training and weights' adjusting watermarking (steps 1 \& 2). 
\end{itemize}



Not only should watermarking techniques be comparable, but so too should the international expectations surrounding them. While the American Executive Order promoting the labelling of synthetic content was rescinded in early 2025 \citep{USA_2025_AI_policy}, global standards for both visible and invisible watermarking of synthetic texts remain under development \citep{C2PA_specification}.
Similarly, new Chinese industrial standards entered into force in September 2025, underscoring the growing challenge of watermarking LLM-generated texts, where discrete symbols cannot embed imperceptible marks as easily as continuous images \citep{kijak2024_image_cryptography_watermarking}. Indeed, the Cyberspace Administration of China now mandates "implicit labelling" for synthetic images and videos, while merely encouraging it for generated text \citep{China_standard_watermark}.

In setting standards for LLM watermarking, it is therefore time for Europe to set out its own framework. Confronted with these emerging global initiatives, the European Union will require robust governance mechanisms to enforce, audit, and certify LLM watermarking practices \citep{Nemecek_2025_interoperability_standards}. Achieving this will first demand a clear operational understanding of what LLM watermarking evaluation entails.

We hope that our interpretation of the European requirements can help shape this understanding, paving the way for future research and empirical assessments of conformity with the AI Act—thus supporting providers’ compliance, guiding Commission auditors, and reinforcing citizens’ trust in AI.  

\section*{Acknowledgments}
Thomas Souverain conducted this research under the the OpenLLM project funded by France-2030. The author is grateful for the precious insights of Dr. Alexei Grinbaum. 

\section*{Competing Interests}
The author declares no competing interests. 


\bibliography{iclr2025_conference}
\bibliographystyle{iclr2025_conference}


\end{document}